\documentclass[twocolumn,preprintnumbers,nofootinbib]{revtex4}
\usepackage{graphicx}
\usepackage{dcolumn}
\usepackage{bm}
\usepackage{amsmath,amssymb,amsfonts}
\usepackage{latexsym}
\usepackage{bbm}
\usepackage{url} 
\usepackage{color}
\begin{document}
\preprint{ACFI-T17-15}

\title{Top Down Electroweak Dipole Operators}

\author{Kaori Fuyuto$^1$}
\email{kfuyuto@umass.edu}
\author{Michael Ramsey-Musolf$^{1,2}$}
\email{mjrm@physics.umass.edu}
\affiliation{$^1$Amherst Center for Fundamental Interactions, Department of Physics,
University of Massachusetts Amherst, MA 01003, USA}
\affiliation{$^2$Kellogg Radiation Laboratory, California Institute of Technology, Pasadena, CA 91125 USA}
\bigskip

\date{\today}

\begin{abstract}
We derive present constraints on, and prospective sensitivity to, the electric dipole moment (EDM) of the top quark ($d_t$) implied by searches for the EDMs of the electron and nucleons. Above the electroweak scale $v$, the $d_t$ arises from two gauge invariant operators generated at a scale $\Lambda \gg v$ that also mix with the light fermion EDMs under renormalization group evolution at two-loop order. Bounds on the EDMs of first generation fermion systems thus imply bounds on $|d_t|$. Working in the leading log-squared approximation, we find that the present upper bound on $|d_t|$ is roughly $10^{-19}$ $e$ cm for $\Lambda = 1$ TeV, except in regions of finely tuned cancellations that allow for $|d_t|$ to be up to fifty times larger. Future $d_e$ and $d_n$ probes may yield an order of magnitude increase in $d_t$ sensitivity, while inclusion of a prospective proton EDM search may lead to an additional increase in reach.

\end{abstract}

\maketitle

\section{Introduction}\label{sec:Intro} 
The search for physics beyond the Standard Model (BSM) lies at the forefront of both high- and low-energy physics.  The properties of the top quark constitute a particularly interesting meeting ground for the two regimes. Theoretically, top quarks may provide a unique window into BSM physics, given that the top Yukawa coupling is large compared to all other Standard Model (SM) fermions. Experimentally, top quarks can be copiously produced in high energy proton-proton collisions, while their indirect effects -- generated via quantum loops -- can be pronounced. Indeed, the breaking of custodial SU(2) symmetry by the top quark-bottom quark mass splitting has a significant impact on the interpretation of electroweak precision tests at the loop level. This sensitivity provided an early handle on the value of the top quark mass and, after the discovery of the top quark, an important test of the self-consistency of the SM at the level of quantum corrections.

The CP properties of top quark interactions is a topic of on-going interest. In the context of electroweak baryogenesis (EWBG) \cite{ewbg}, CP-violating (CPV) interactions of the top quark  with an extended scalar sector can yield the observed cosmic baryon asymmetry \cite{Tulin:2011wi,Cline:2011mm,Jiang:2015cwa,Huang:2015izx,Kobakhidze:2015xlz,Fuyuto:2017ewj}. 
The presence of BSM CPV in the top quark sector may also appear in the guise of a top electric dipole moments (EDM) and chromo-electric dipole moment (CEDM), two of a number of possible higher dimension top quark operators. Since the top (C)EDM is chirality changing, it can be significantly enhanced compared to light fermion (C)EDMs by the large top Yukawa coupling.

While direct collider probes of the (C)EDM have  been studied extensively \cite{Gupta:2009wu,Hayreter:2015ryk,Bernreuther:2015yna,Rindani:2015vya,Hayreter:2014hha,Hioki:2013hva,Bernreuther:2013aga,Hayreter:2013kba,Baumgart:2012ay,Biswal:2012dr,Hioki:2012vn,Choudhury:2012np,Baur:2004uw, Baur:2006ck, Bouzas:2012av, Fael:2013ira, Fayazbakhsh:2015xba, Etesami:2016rwu, Aguilar-Saavedra:2014iga, Chien:2015xha}, a complementary way to access the top EDM ($d_t$) and CEDM (${\tilde d}_t$) is through  their indirect effects, such as the resulting, radiatively-induced light fermion EDMs.
This possibility has been explored in several studies \cite{CorderoCid:2007uc, Kamenik:2011dk, Cirigliano:2016njn, Cirigliano:2016nyn}. The most powerful limit on $d_t$ appears to result from the limit on the EDM of the electron $|d_e|<8.7\times 10^{-29}~e~{\rm cm}~(90\%~{\rm C.L.})$ \cite{Baron:2013eja} (see also the recent result using HfF$^+$, $|d_e|<1.3\times 10^{-28}~e~{\rm cm}~(90\%~{\rm C.L.})~ $\cite{Cairncross:2017fip}), implying $|d_t|<5.0\times 10^{-20}~e~{\rm cm}$ ($90\%~{\rm C.L.}$) \cite{Cirigliano:2016njn, Cirigliano:2016nyn}.

In this study, we focus on $d_t$. If it is generated by BSM physics at a scale $\Lambda$ that lies well above the electroweak scale $v=246$ GeV, then it is likely that two dimension-six CPV dipole operators emerge, coupling respectively to the U(1)$_Y$ and SU(2)$_L$ gauge bosons. We henceforth denote these operators as ${\cal O}_{tB}$ and ${\cal O}_{tW}$, respectively. Denoting their coefficients as $C_{tB(W)}/\Lambda^2$, we note that the presence of CPV implies that the dimensionless Wilson coefficients $C_{tB(W)}$ are, in general, complex.  After electroweak symmetry breaking (EWSB), one linear combination yields $d_t$ at tree-level. The operators ${\cal O}_{tB}$ and ${\cal O}_{tW}$ will also radiatively generate all other light fermion EDMs at two-loop order. Bounds on $d_e$ as well as on the neutron EDM, $d_n$, then yield (in principle) complementary constraints on  $C_{tB(W)}$, with corresponding implications for $d_t$. 

In what follows, we perform an explicit two-loop computation of the light fermion EDMs induced by ${\cal O}_{tB(W)}$, retaining the leading $\ln^2 (\Lambda/v)$ contributions. After translating the light quark EDMs into $d_n$, we derive constraints on the $C_{tB(W)}/\Lambda^2$, along with the corresponding implications for $d_t$, using the present neutron and electron EDM bounds. We will make no {\em a priori} assumptions about the relationships between the $C_{tB}$ and $C_{tW}$ at the scale $\Lambda$, endeavoring to be as model-independent as possible. In these respects, our analysis complements the earlier studies in Refs.~\cite{CorderoCid:2007uc, Kamenik:2011dk, Cirigliano:2016njn, Cirigliano:2016nyn}. In this context, we also find that there exist regions where cancellations between these two operators can considerably weaken the generic constraints, albeit with some degree of fine-tuning. Looking ahead, we illustrate the potential reach of next generation electron and nucleon EDM searches.

\section{Effective operators}\label{sec:effective_operator}
\begin{figure}[t]
\begin{center}
\begin{tabular}{cc}
\includegraphics[width=3cm]{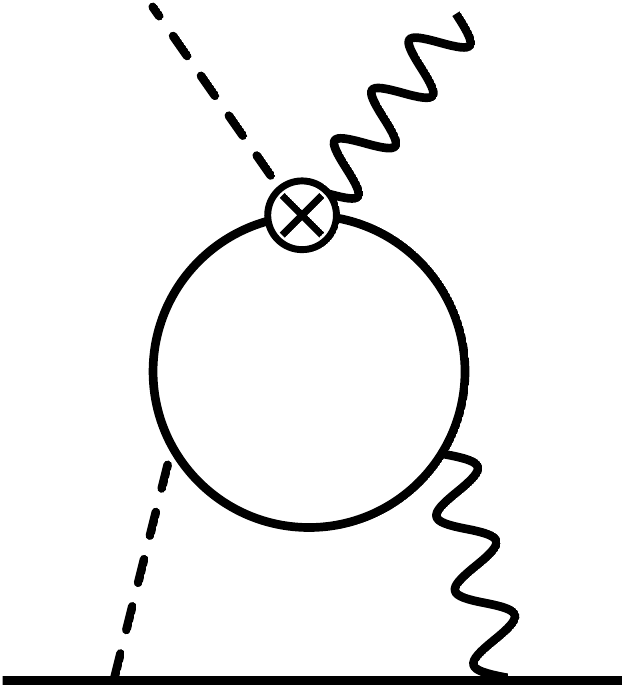} 
\includegraphics[width=4.7cm]{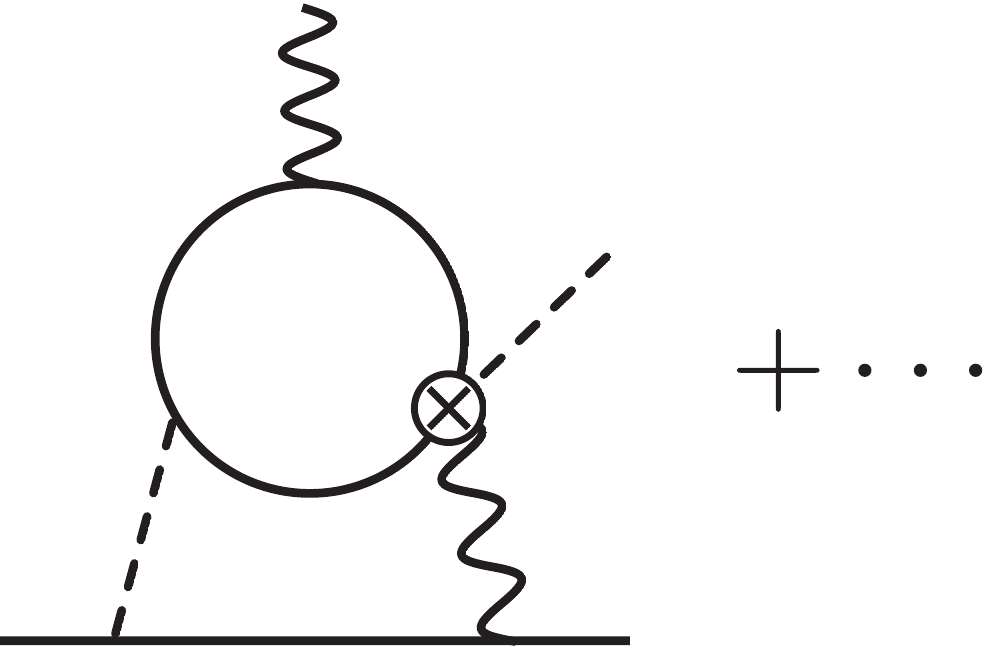} 
\end{tabular}
\begin{tabular}{cc}
\includegraphics[width=3cm]{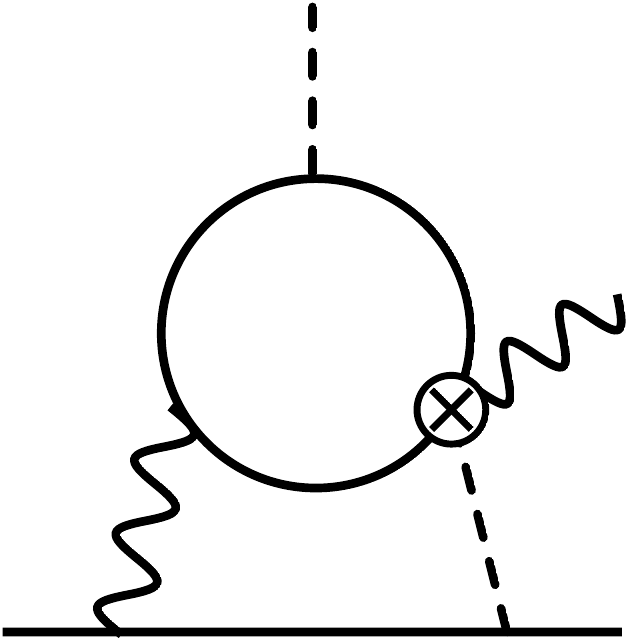} 
\includegraphics[width=4.5cm]{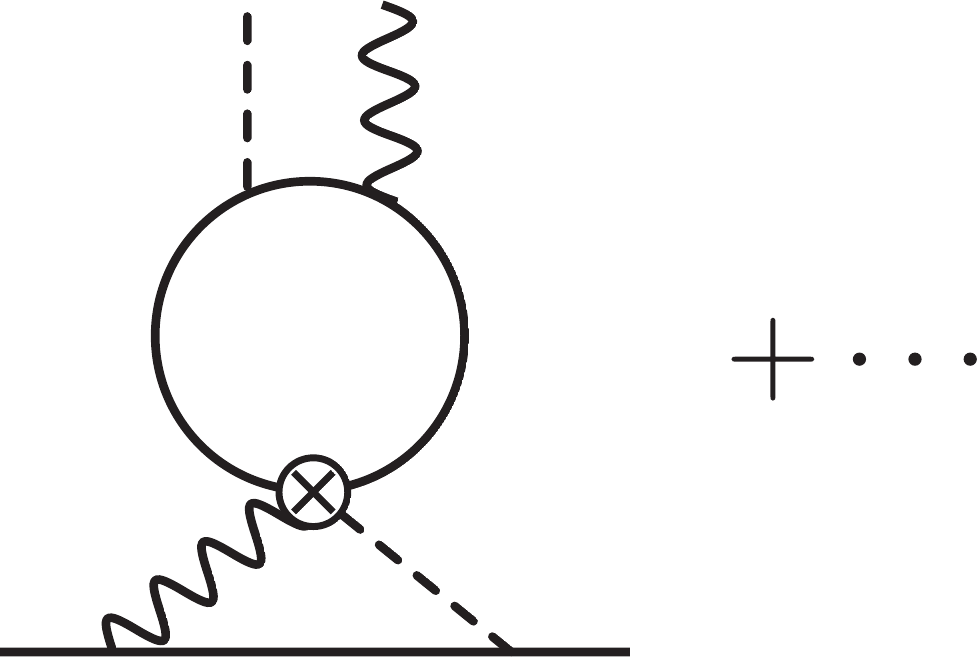}
\end{tabular}
\end{center}
\caption{The Barr-Zee diagrams induced by the dipole operator of the top quark. The circled cross mark denotes the top quark dipole operator, and the other wavy lines correspond to the gauge fields $B$ or $W^A$. While the upper two diagrams lead to the dipole operator of the up quark, the lower diagrams yield those of the electron and down quark. The $``+\cdots"$ indicate additional topologies that contribute to the light fermion EDMs. }
\label{fig:bar_zee}
\end{figure}
To set the conventions for our analysis, we start with the CPV effective Lagrangian generated by BSM physics at the scale $\Lambda$ \cite{Cirigliano:2016njn,Cirigliano:2016nyn}:
\begin{align}
{\cal L}_{\rm eff}=&-\frac{1}{\Lambda^2}\sum_{f=e,u,d,t}\left(\frac{g_1}{\sqrt{2}}{C_{fB}}{\cal O}_{fB}+\frac{g_2}{\sqrt{2}}C_{fW}{\cal O}_{fW}+{\rm h.c.}\right)\nonumber\\
&+\frac{1}{\Lambda^2}\sum_{X=B,W}C_{H\tilde X}{\cal O}_{H\tilde X}\nonumber\\
&+\frac{1}{\Lambda^2}\sum_{F=L,Q,f=e,d,t}\left(C^{(i)}_{FfF'f'}{\cal O}_{FfF'f'}^{(i)}+{\rm h.c.}\right),  \label{effective_lag} 
\end{align}
where the first line indicates the dipole operators 
\begin{align}
{\cal O}_{eB}&=\bar{L}\sigma^{\mu\nu}e_RHB_{\mu\nu},\nonumber\\
{\cal O}_{eW}&=\bar{L}\sigma^{\mu\nu}e_R\tau^AHW^{A}_{\mu\nu},\nonumber\\
{\cal O}_{tB}&=\bar{Q}\sigma^{\mu\nu}t_R\tilde{H}B_{\mu\nu},\nonumber\\
{\cal O}_{tW}&=\bar{Q}\sigma^{\mu\nu}t_R\tau^A\tilde{H}W^{A}_{\mu\nu} \label{effective_operator}.
\end{align}
The second and third lines represent gauge-Higgs and 4-fermi operators
\begin{align}
{\cal O}_{H\tilde B}&=g^2_1H^{\dagger}H\tilde B_{\mu\nu}B^{\mu\nu},\nonumber\\
{\cal O}_{H\tilde W}&=g^2_2H^{\dagger}H\tilde W^A_{\mu\nu}W^{A\mu\nu},\nonumber\\
{\cal O}_{H\tilde WB}&=g_1g_2H^{\dagger}\tau^AH\tilde W^A_{\mu\nu}B^{\mu\nu},
\end{align}
and
\begin{align}
{\cal O}_{\ell e qt}^{(3)}&=(\bar{L}^a\sigma^{\mu\nu}e_R)\epsilon_{ab}(\bar{Q}^b\sigma_{\mu\nu}t_R),\nonumber\\
{\cal O}_{qtqd}^{(1)}&=(\bar{Q}^at_R)\epsilon_{ab}(\bar{Q}^bd_R),\nonumber\\
{\cal O}_{qtqd}^{(8)}&=(\bar{Q}^a\tau^At_R)\epsilon_{ab}(\bar{Q}^b\tau^Ad_R).
\end{align}
Here, $L$ and $Q$ are the lepton and quark doublets, $e_R~(t_R)$ is the right-handed electron (top quark), $\tau^A$ is the Pauli matrix, and $H$ is the Higgs doublet with $\tilde{H}=i\tau^2H^*$;  $B_{\mu\nu}$ and $W^{A}_{\mu\nu}$ are the U(1)$_Y$ and SU(2)$_L$ field strengths, respectively; and  $g_1$ and $g_2$ represent their gauge couplings; $\tilde X$ is defined as $\epsilon_{\mu\nu\alpha\beta}X^{\alpha\beta}$; $a$ and $b$ are the SU(2)$_L$ indices.
The dipole operators for the up (down) quark  ${\cal O}_{uB,uW}~({\cal O}_{dB,dW})$ are also given by the same structure as ${\cal O}_{tB,tW}~({\cal O}_{eB,eW})$.
For a listing of the complete set of dimension-six CPV operators, see, {\em e.g.}, \cite{Buchmuller:1985jz, Grzadkowski:2010es}.

After EWSB, the dipole operators in Eq. (\ref{effective_lag}) produce the EDMs 
\begin{align}
{\cal L}_{\rm eff}\ni -\frac{i}{2}\sum_{f=e,u,d,t}d_f\bar{f}\sigma^{\mu\nu}\gamma_5fF_{\mu\nu},
\end{align}
with $F_{\mu\nu}$ being the photon field strength tensor. The coupling $d_f$ is related to the Wilson coefficients of the operators
\begin{align}
d_{e(d)}&=\frac{ev}{\Lambda^2}\left\{{\rm Im}({C}_{e(d)B})-{\rm Im}({C}_{e(d)W}) \right\},\nonumber\\
d_{t(u)}&=\frac{ev}{\Lambda^2}\left\{{\rm Im}({C}_{t(u)B})+{\rm Im}({C}_{t(u)W}) \right\} \label{EDMs}.
\end{align}
The opposite relative sign between the $C_{fB}$ and $C_{fW}$ for up- and down-type fermions is due to their isospin projection quantum numbers. To facilitate comparison with the experimental EDM limits, it is useful to express a factor of $ev/\Lambda^2$ with units of fm\footnote{Since our definitions of the dipole operators are accompanied with a factor of $1/\sqrt{2}$, the coefficient of $ev/\Lambda^2$ becomes smaller that in \cite{Engel:2013lsa}.}
\begin{align}
\frac{ev}{\Lambda^2}=\frac{e}{v}\left(\frac{v}{\Lambda}\right)^2\simeq (7.8\times 10^{-4}~e~{\rm fm})\left(\frac{v}{\Lambda}\right)^2.
\end{align}

In addition to the bounds on $|d_e|$ quoted above\footnote{The limit is obtained by assuming that the ThO EDM does not receive a contribution from semileptonic four-fermion interactions.}, we consider the constrains implied by the light-quark contributions to $d_n$\footnote{Although the EDM of the strange quark and chromo EDMs also contribute to the neutron EDM, we do not include them, here.}, whose experimental limit is $|d_n|<3.0\times 10^{-26}~e~{\rm cm}~(90\%~{\rm C.L.})$ \cite{Afach:2015sja}. 
As we discuss below, the $d_e$-contributions from ${\cal O}_{tB}$ and ${\cal O}_{tW}$ may cancel in some finely-tuned portions of parameter space. Inclusion of the $d_n$ constraints may provide a complementary probe of this \lq\lq cancellation region". Outside of this region, present EDM limits imply an upper bound on $|d_t|\lesssim 10^{-19}~e~{\rm cm}$, depending on the value of $\Lambda$.  Looking to the future, next generation EDM searches may reach the levels of sensitivity: $|d_e|<1.0\times 10^{-29}~e~{\rm cm}$ and $|d_n|<3.0\times 10^{-28}~e~{\rm cm}$ \cite{NSAC2015}, implying an order of magnitude increase in the sensitivity to $d_t$.  In addition, efforts are underway to develop storage ring proton EDM search with sensitivity $10^{-29}~e~{\rm cm}$ \cite{Kumar:2013qya}. For the scenario considered here, the constraints from diamagnetic atom EDM searches, such as that of the $^{199}$Hg atom \cite{Graner:2016ses} can be comparable to those from $d_n$. Although the latest $^{199}$Hg result yields an upper bound on $|d_n|$ that is roughly two times stronger than the direct limit, we expect the latter to become considerably more stringent with the next generation experiments. Consequently, we will use the direct $d_n$ bounds in what follows. 

\section{Loop calculations}\label{sec:loop}

The existence of the top quark dipole operators in Eq. (\ref{effective_lag}) at a renormalization scale $\mu=\Lambda$ will lead to non-vanishing electron and light-quark dipole operators through the two-loop Barr-Zee diagrams of Fig.~\ref{fig:bar_zee}. This effect corresponds to the electroweak operator mixing in the renormalization group evolution (RGE) from $\Lambda$ to $v$,  
thereby relating the Wilson coefficients of the electron and light quark dipole operators at the EW scale  to $C_{tB}(\Lambda)$ and $C_{tW}(\Lambda)$. Below the scale $v$, we integrate out the heavy SM degrees of freedom ($t$, $W$, $Z$, and $h$), and the dominant contributions when running to the low-energy scale relevant to experiment involve SU(3)$_C$ interactions.  The upper two diagrams induce the up quark EDM, the lower two diagrams yield the electron and down quark EDMs. This assignment can be understood by considering which Higgs field is chosen as an external particle. Each diagram has two opposite fermion flows (corresponding to distinct Wick contractions), as well as topologies involving crossing of the scalar and gauge boson lines.

\begin{figure}[t]
\begin{center}
\begin{tabular}{cc}
\includegraphics[width=3cm]{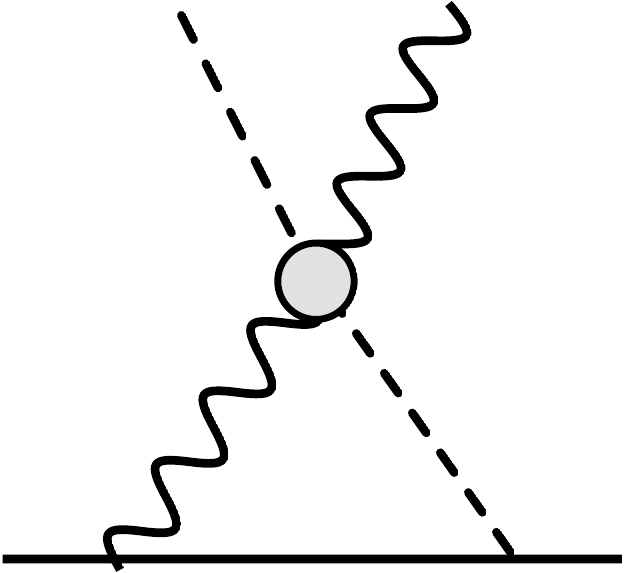} 
\includegraphics[width=3cm]{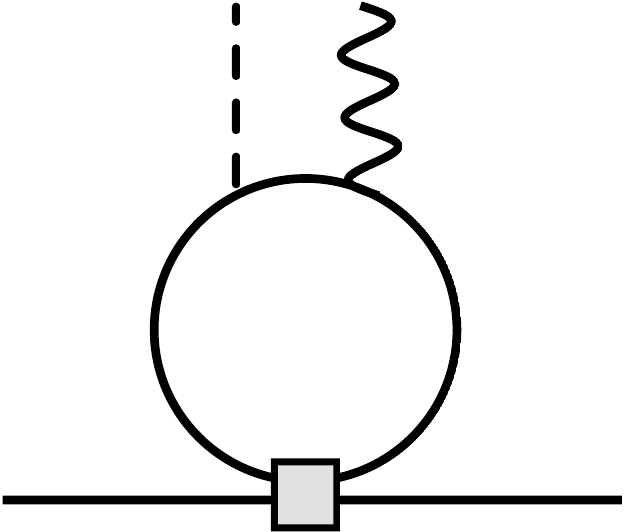} 
\end{tabular}
\end{center}
\caption{The one-loop diagrams with the counter terms for the upper and lower loops in the Barr-Zee diagrams of Fig. \ref{fig:bar_zee}. The shaded circle and square imply their counter terms.}
\label{fig:bar_zee_CT}
\end{figure}

In addition to the overall logarithmic divergence associated with these diagrams, logarithmically divergent one-loop sub-graphs associated with the upper and lower loops in Fig.~\ref{fig:bar_zee} correspond to mixing between  ${\cal O}_{tB,W}$ and $\mathcal{O}_{H{\tilde B}, {\tilde W}, {\tilde W}B}$ and ${\cal O}^{(3, 1,8)}_{\ell e qt, qtqd}$, respectively. Consequently, one must include the counter terms associated with these operators, as shown in Fig.~\ref{fig:bar_zee_CT}. We note that the right diagram in Fig. \ref{fig:bar_zee_CT} results from only the subgraph in the lower right diagram of Fig. \ref{fig:bar_zee}, because only this subgraph has a divergence. 

We perform the computation using dimensional regularization in $d=4-\epsilon$ dimensions and renormalization in the minimal subtraction (MS) scheme. For purposes of this analysis, wherein we seek to obtain the order of magnitude constraints on $C_{tB(W)}(\Lambda)$, it is useful to observe that the EW running yields an enhancement factor of $\ln^2(\Lambda/v)$, as well as sub-dominant $\ln(\Lambda/v)$ terms.  The anomalous dimension associated with the latter is renormalization scheme-dependent and introduce an additional dependence on the Wilson coefficients $C_{H{\tilde W}}$, {\em etc.}. Here, we retain only the leading $\ln^2$ contributions, deferring a treatment of the sub-leading log terms to a future publication \cite{KF_MJRM}.  
(For analogous $\ln^2$ contributions in other contexts, see, {\em e.g.} Refs.~\cite{RamseyMusolf:1999nk,Knecht:2001qg,RamseyMusolf:2002cy,Hisano:2012cc}.)

In the leading $\ln^2$-approximation, the resulting Wilson coefficients for the light fermion ($f=e,u,d$) dipole operators at the scale $v$ are  
\begin{align}
C_{fB}(v)&= -\frac{1}{2}\left( A_{f}C_{tB}+B_{f}C_{tW}\right)\ln^2\left(\frac{\Lambda}{v}\right),\nonumber\\
C_{fW}(v)&=-\frac{1}{2}\left( D_{f}C_{tB}+E_fC_{tW}\right)\ln^2\left(\frac{\Lambda}{v}\right),
\label{LO_results}
\end{align}
where we assume that  $C_{fB,fW}(\Lambda)=0$. The coefficients of $A_f,~B_f,~D_f$ and $E_f$ for $f=e$ and $d$ are given by
\begin{align}
A_{f}&={\cal Y}_{f}\bigg[-6\left(Y_{F}+Y_{f}\right)\left(Y_Q+Y_t\right)g^2_1+\frac{3}{2}g^2_2 \bigg],\nonumber\\
B_{f}&={\cal Y}_{f}~6\left(Y_Q+Y_t\right)g^2_2,\nonumber\\
D_{f}&={\cal Y}_{f}~2\left(Y_{F}+Y_{f}\right)g^2_1,\nonumber\\
E_{f}&={\cal Y}_{f}\bigg[2\left(Y_{F} +Y_{f}\right)\left(Y_Q+Y_t\right)g^2_1-\frac{5}{2}g^2_2 \bigg],
\end{align}
where $F=L$ or $Q$ for $f=e$ or $d$. These of the up quark are given by
\begin{align}
A_u&=-{\cal Y}_{u}\bigg[4\left(Y_{Q}+Y_u\right)\left(Y_Q+Y_t\right)g^2_1+\frac{3}{2}g^2_2 \bigg],\nonumber\\
B_u&=-{\cal Y}_{u}~3\left(Y_Q+Y_t\right)g^2_2,\nonumber\\
D_u&=-{\cal Y}_{u}\left(Y_{Q}+Y_{t}\right)g^2_1,\nonumber\\
E_u&=-{\cal Y}_{u}\bigg[2\left(Y_{Q} +Y_{u}\right)\left(Y_Q+Y_t\right)g^2_1+g^2_2 \bigg],
\end{align}
where ${\cal Y}_{f}=N_Cy_{f}y_t/(4\pi)^4$ with $N_C=3$ and the hyper charges $Y_L=-1/2,~Y_e=-1,~Y_Q=1/6,~Y_{t(u)}=2/3$ and $Y_d=-1/3$. ${\cal Y}_{e}$ is roughly an order of magnitude smaller than ${\cal Y}_{u,d}$ due to the Yukawa coupling. 

Using these results, it is straightforward to obtain the  light fermion EDMs as defined in Eq. (\ref{EDMs}):
\begin{align}
&d_{e(d)}=-\frac{e}{2v}\left(\frac{v}{\Lambda}\right)^2\ln^2\left(\frac{\Lambda}{v}\right)\nonumber\\
&\times
\bigg[\left(A_{e(d)}-D_{e(d)}\right){\rm Im}(C_{tB})+\left(B_{e(d)}-E_{e(d)}\right){\rm Im}(C_{tW}) \bigg]\nonumber \\
&d_{u}=-\frac{e}{2v}\left(\frac{v}{\Lambda}\right)^2 \ln^2\left(\frac{\Lambda}{v}\right)\nonumber\\
&\times
\bigg[\left(A_{u}+D_{u}\right){\rm Im}(C_{tB})+\left(B_{u}+E_{u}\right){\rm Im}(C_{tW}) \bigg].
\end{align}

\begin{figure}[t]
\begin{center}
\includegraphics[width=6cm]{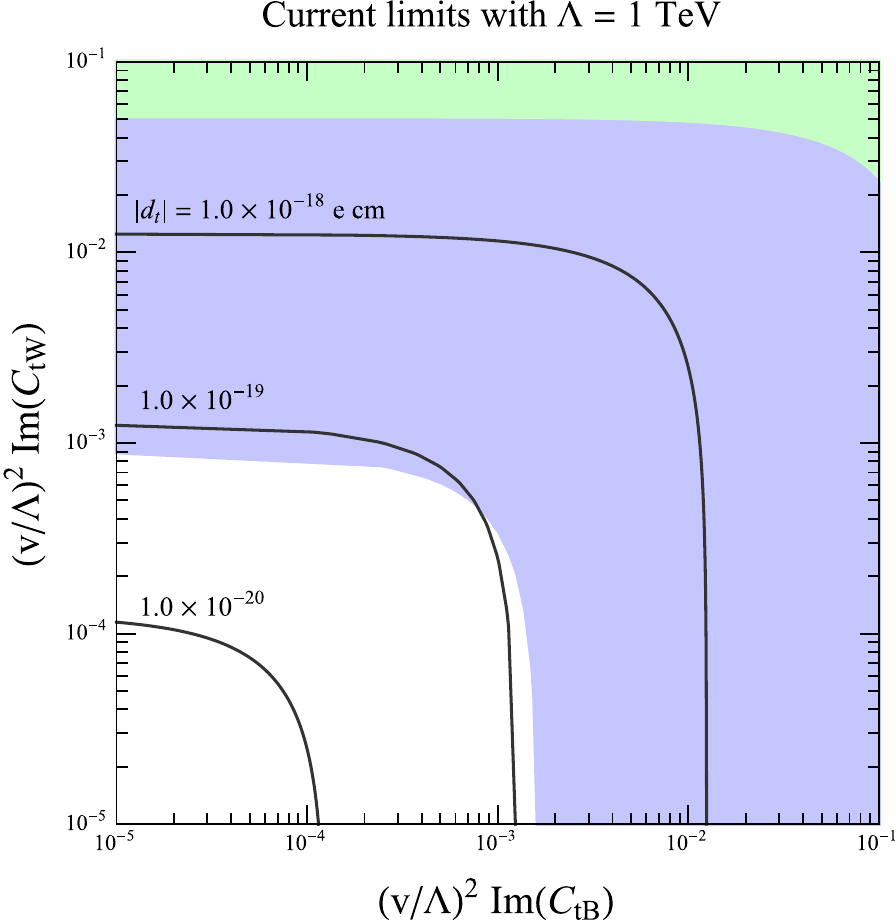}
\includegraphics[width=6cm]{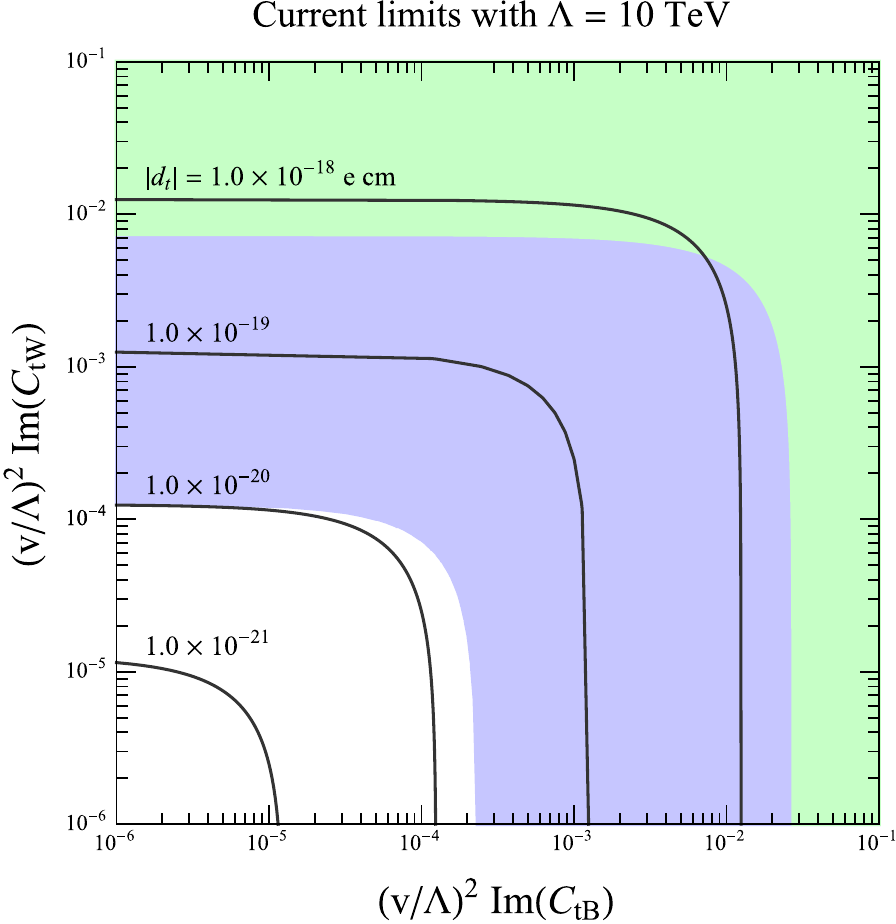}   
\end{center}
\caption{Excluded regions of $(v/\Lambda)^2{\rm Im}(C_{tB})$ and $(v/\Lambda)^2{\rm Im}(C_{tW})$ by the EDMs of the electron (blue) and neutron (green). The new physics scale is taken at $1~(10)$ TeV in the upper (lower) figure. The black lines are the top EDM, $|d_t|=1.0\times 10^{-18},~10^{-19}~10^{-20}$ and $10^{-21}~e~{\rm cm}$, from top to bottom. }
\label{fig:result_po}
\end{figure}
\begin{figure}[t]
\begin{center}
\includegraphics[width=6cm]{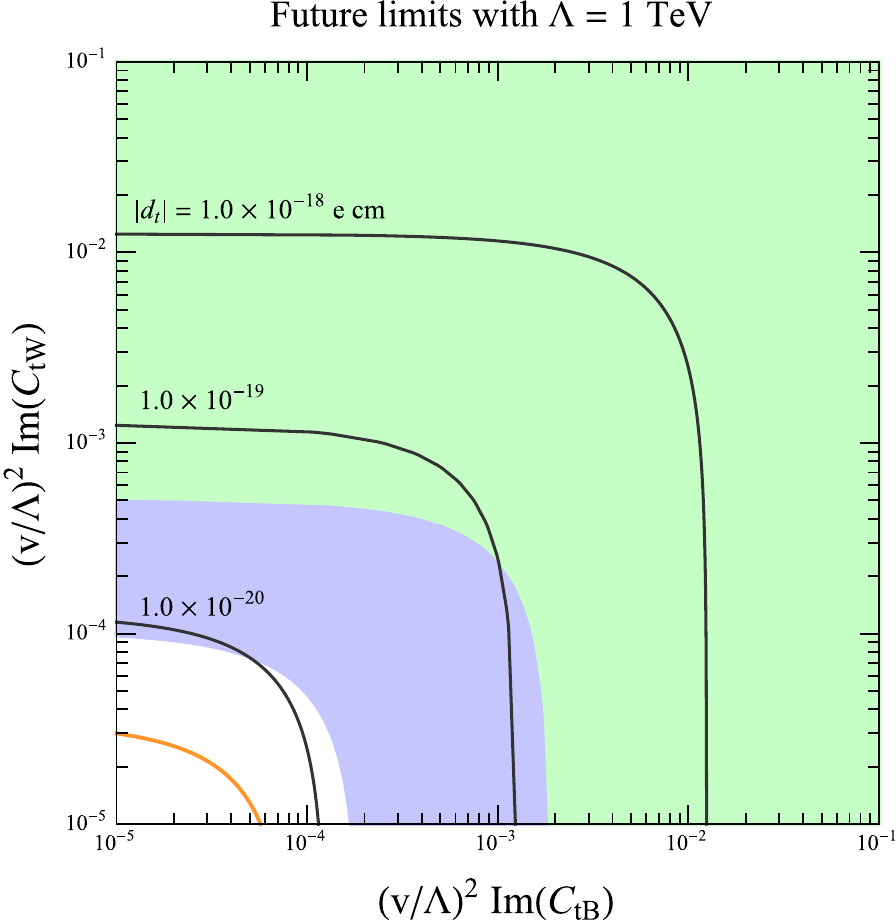} 
\includegraphics[width=6cm]{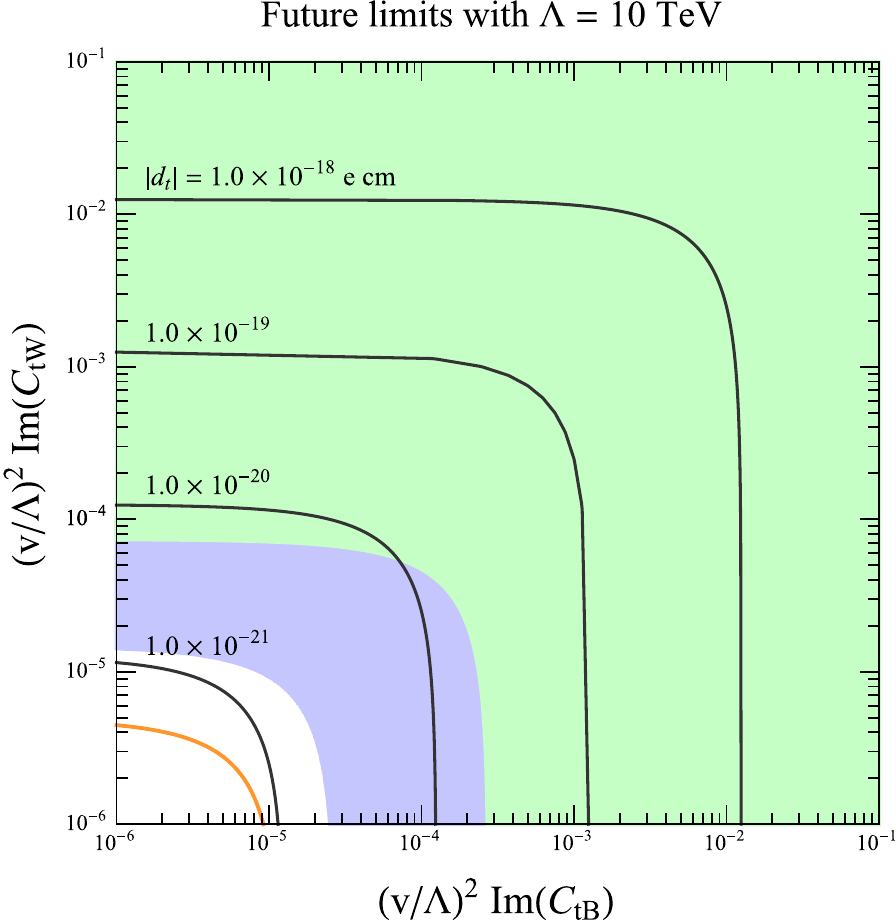} 
\end{center}
\caption{Excluded regions with the future sensitivities of $|d_n|=3.0\times 10^{-28}~e~{\rm cm}$ (green) and $|d_e|=1.0\times 10^{-29}~e~{\rm cm}$ (blue).
The upper (lower) figure takes $\Lambda=1~(10)$ TeV.
The orange line represents the proton EDM of $|d_p|=1.0\times 10^{-29}~e~{\rm cm}$.}
\label{fig:result_po_future}
\end{figure}

In general, the $d_f$ depend more strongly on Im$(C_{tW})$ than on Im$(C_{tB})$, a feature due in part to the dependence on $g_2$.  Specifically, the Im$(C_{tB})$ contribution depends on $g^2_2$ comes from only $A_f$, while both $B_f$ and $E_f$ contain $g^2_2$ contributions. The dependence on $\Lambda$ comes from $(v/\Lambda)^2$ and $\log^2(\Lambda/v)$ factors. When translating the limits on $d_{e(n)}$ into bounds on $|d_t|$, the $(v/\Lambda)^2$-dependence that is common to all EDMs. To assess the impact of the remaining logarithmic dependence, in our numerical analyses we consider two benchmark choices: $\Lambda = 1$ and $10$ TeV.  The ratio $\ln^2(\Lambda=1~{\rm TeV}/v)/\ln^2(\Lambda=10~{\rm TeV}/v)$ is about $0.14$.

For the light quark EDMs, we take into account the QCD contributions to their evolution from the EW scale to the low-energy scale \cite{Hisano:2012cc,Shifman:1976de,Ciuchini:1993fk, Degrassi:2005zd, Dekens:2013zca, Fuyuto:2013gla}. As clearly discussed in \cite{Degrassi:2005zd}, the effect suppresses the dipole operators at the low-energy. We choose the low-energy scale $\Lambda_\mathrm{had} = 2$ GeV in order to match onto the lattice QCD computation of the resulting neutron EDM given in~\cite{Bhattacharya:2015esa, Bhattacharya:2015wna}. We obtain $d_q(\Lambda_\mathrm{had}) = 0.85 d_q(v)$.

\section{Results}\label{sec:results}
\begin{figure}[t]
\begin{center}
\includegraphics[width=6cm]{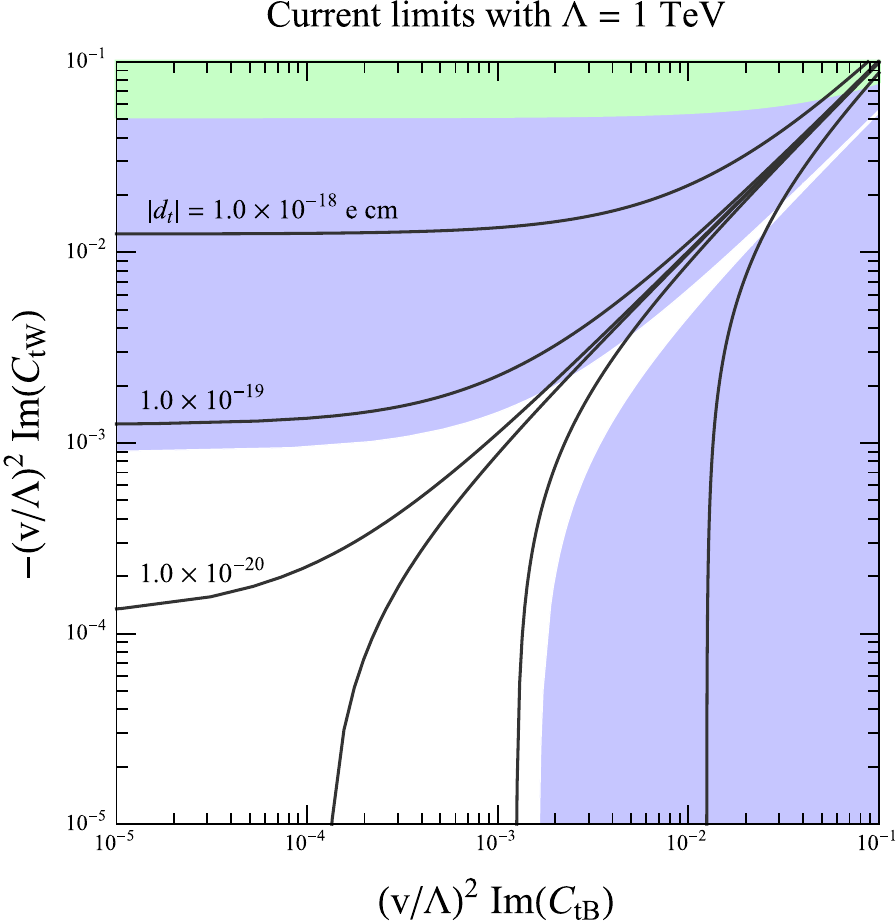}
\includegraphics[width=6cm]{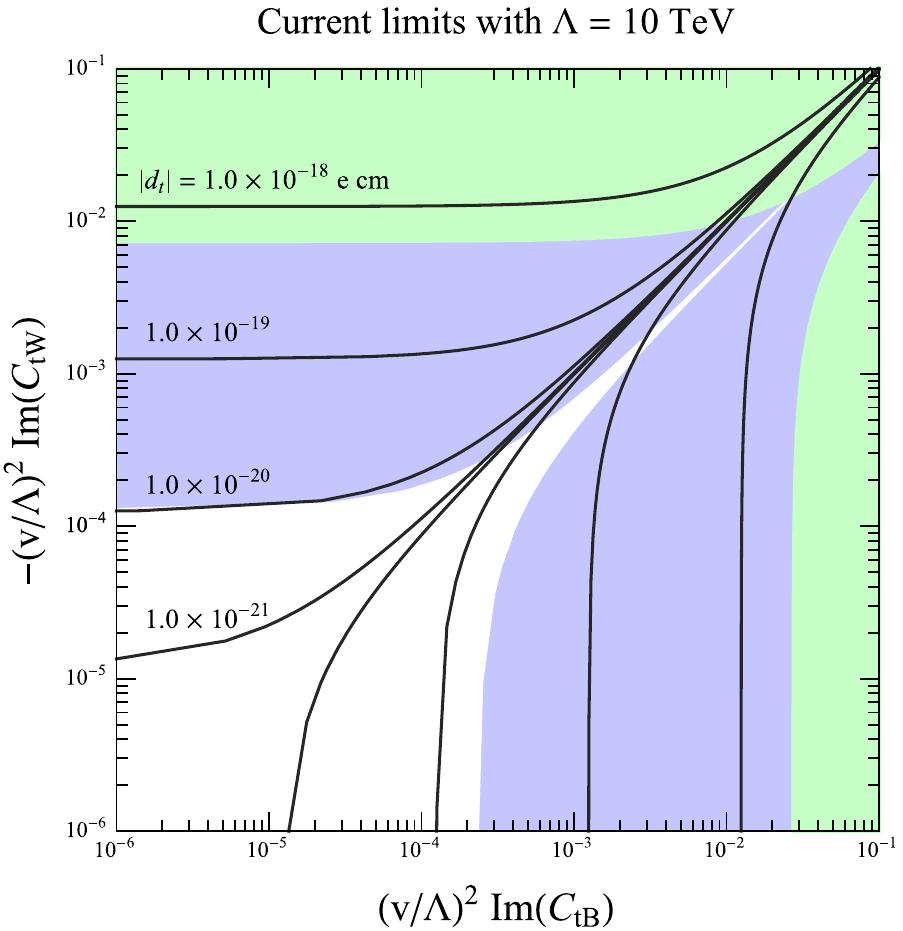}   
\end{center}
\caption{Excluded regions of $(v/\Lambda)^2{\rm Im}(C_{tB})$ and $-(v/\Lambda)^2{\rm Im}(C_{tW})$ by the electron (blue) and neutron (green) EDMs.  The upper (lower) figure takes $\Lambda=1~(10)$ TeV.}
\label{fig:result_ne}
\end{figure}
\begin{figure}[t]
\begin{center}
\includegraphics[width=6cm]{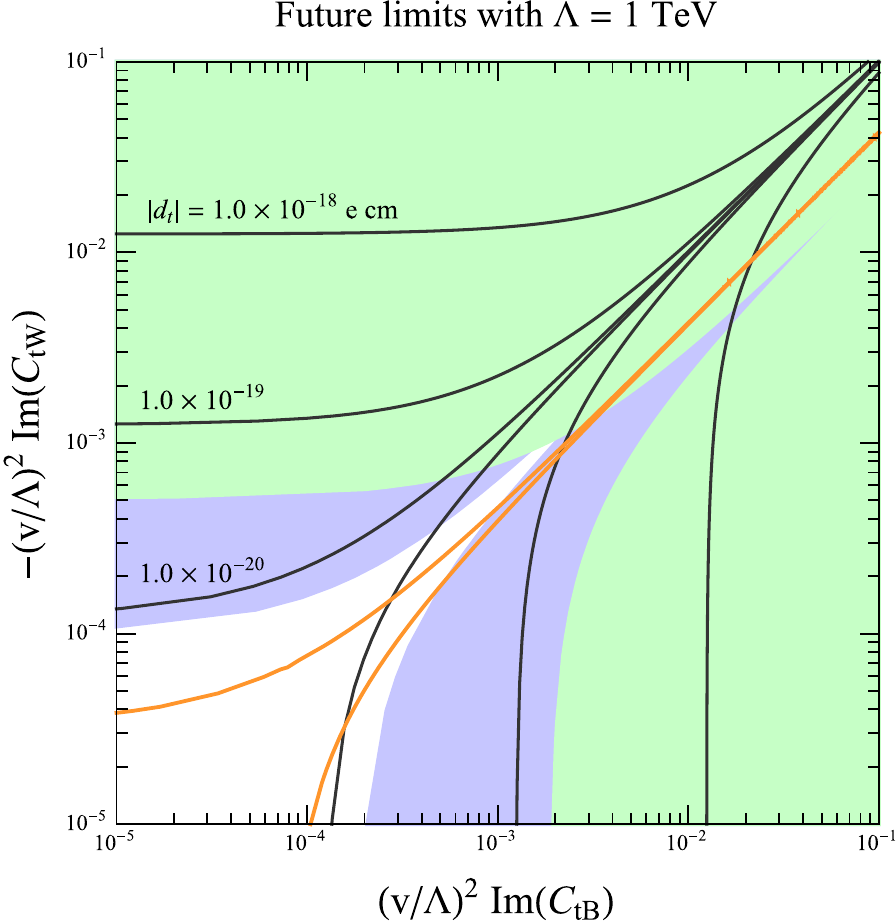} 
\includegraphics[width=6cm]{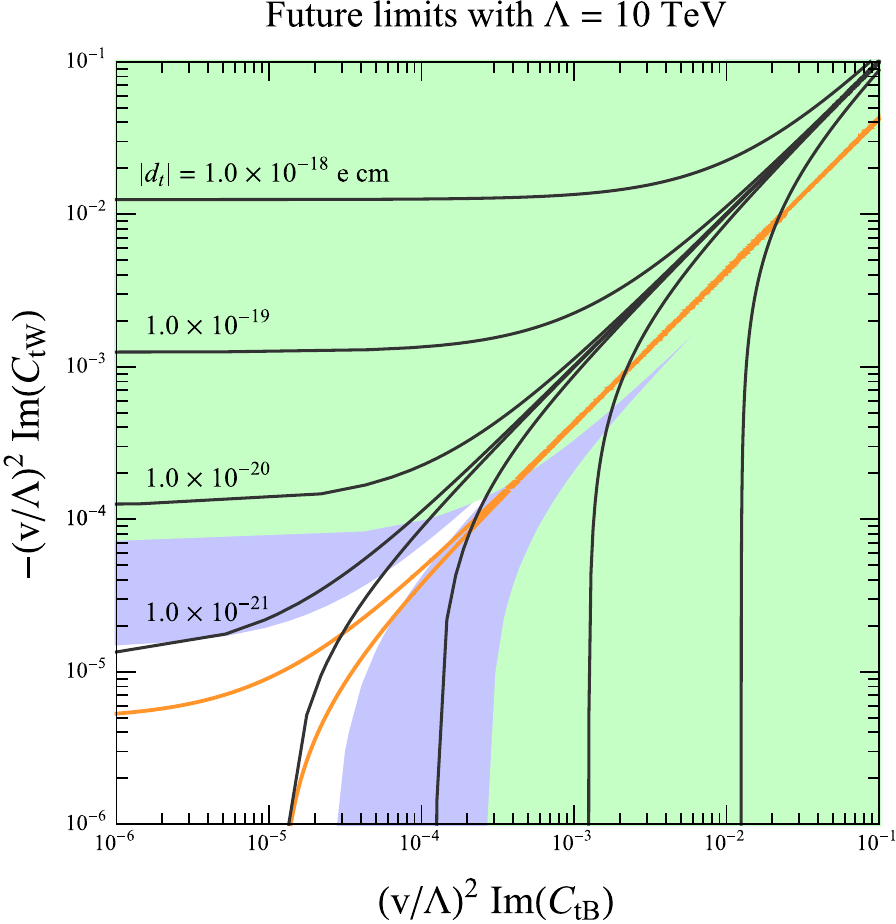} 
\end{center}
\caption{Excluded regions by the EDMs of the electron (blue) and neutron (green) with the future sensitivities. The upper (lower) figure takes $\Lambda=1~(10)$ TeV. The proton EDM of $|d_p|=1.0\times 10^{-29}~e~{\rm cm}$ is drawn by the orange line.}
\label{fig:result_ne_future}
\end{figure}

It is useful to consider the constraints on $(v/\Lambda)^2 {\rm Im}(C_{tB(W)})$ since the EDM definitions absorb the leading $(v/\Lambda)^2$ factor is noted above. The present and prospective bounds are shown in Figs.~\ref{fig:result_po}-\ref{fig:result_ne_future}. In addition to considering the two benchmark choices for $\Lambda$, we also consider two cases, corresponding to  ${\rm Im}(C_{tB})$ and ${\rm Im}(C_{tW})$ having the same (positive) sign  or opposite sign. The latter exhibits the possibility of finely-tuned cancellations.

Figure \ref{fig:result_po} shows the present constraints for the same sign case for the two different benchmark choices for $\Lambda$.  The blue and green shaded regions are excluded by the limits in $d_e$ and $d_n$, respectively. The black contours represent values of constant top quark EDM. 
For $\Lambda=1$ (10) TeV, we find that  $|d_t| \lesssim 1.3\times 10^{-19}~ (1.8\times 10^{-20})~e$ cm in the limit of ${\rm Im}(C_{tW})=0$. Note that the maximum value for $\Lambda = 10$ TeV is roughly 0.14 times smaller than for $\Lambda = 1$ TeV, as expected from the $\ln^2$ dependence on $(v/\Lambda)^2$. We observe that our upper bound for $\Lambda= 1$ TeV is somewhat larger than obtained by the authors of Ref.~\cite{Cirigliano:2016njn,Cirigliano:2016nyn}, who assumed in their numerical analysis that only one linear combination of ${\rm Im}(C_{tB})$ and ${\rm Im}(C_{tW})$ corresponding to non-vanishing $d_t$ exists at the scale $\Lambda$. Although the computation in Refs.~\cite{Cirigliano:2016njn, Cirigliano:2016nyn} was performed using the broken phase basis, we speculate that the difference in our limits in $d_t$ results primarily from the assumptions about the ${\rm Im}(C_{tB(W)})(\Lambda)$ . 

The prospective impact of future EDM searches is illustrated in Fig.~\ref{fig:result_po_future}, where we assume 90\% C.L. limits  of $|d_n|=3.0\times 10^{-28}~e~{\rm cm}$ and $|d_e|=1.0\times 10^{-29}~e~{\rm cm}$.  For the same sign case, we see that the prospective constraint from $d_e$ would still be stronger than from $d_n$. Na\"ively, one would expect the impact of future experiments with these sensitivities to be comparable, since the light fermion EDMs scale linear with the fermion masses and the ratio of the light quark and electron EDMs is roughly a factor of ten. The somewhat stronger $d_e$ sensitivity results from a factor of $3$ difference in the future sensitivities and the suppression of the light quark EDMs due to the QCD evolution from the weak to hadronic scales. The resulting prospective bound on $d_t$ for $\Lambda = 1$ (10) TeV is $|d_t|\lesssim1.5\times 10^{-20}~(2.1\times10^{-21})~e$ cm.
We also include the possibility of a future proton EDM search, with sensitivity $|d_p|=1.0\times 10^{-29}~e~$cm indicated by the orange contour. Should a search with this sensitivity be realized, a top quark EDM of order $10^{-20~(21)}$ for $\Lambda = 1$ (10) TeV could be probed.

Next, we consider the opposite sign case, with present and prospective constraints indicated in Figs.~\ref{fig:result_ne} and \ref{fig:result_ne_future}, respectively. Here, the situation is more subtle than for the same sign case, as there exist regions where cancellations between ${\rm Im}(C_{tB})$ and ${\rm Im}(C_{tW})$ can lead to the absence of any constraint from $d_e$. The present $d_n$ bounds are not yet sufficiently strong to probe this \lq\lq cancellation region" for $d_t\lesssim 10^{-18~(19)}~e$ cm for $\Lambda=1~(10)$ TeV.  Although the existence of this loophole admittedly requires a degree of fine tuning, a similar possibility of canceling contributions has been noted elsewhere in the case of the minimal supersymmetric SM and proposed as a possible solution to the  \lq\lq SUSY CP problem" \cite{Ibrahim:1997gj, Ibrahim:1998je, Falk:1998pu, Brhlik:1998zn}. Outside of this region, the present upper bound on $d_t$ is the same as for the same sign case. As seen in Fig.~\ref{fig:result_ne_future}, the future bound of $d_n$ closes the loophole and yields of $|d_t|\lesssim1.0\times 10^{-19~(20)}~e$ cm for $\Lambda=1~(10)$ TeV.  On the other hand, the electron EDM with the future sensitivity plays a complementary role that covers the region where $|d_n|=0$.  The prospective, future proton EDM experiment gives a sensitivity to $d_t$ with a similar order of magnitude, perhaps increasing the reach by  factor of two.
We summarize the present and future limits on $d_t$ in Table \ref{talbe:limits_dt} and \ref{talbe:limits_dt_loophole}.
\begin{table}[t]
\caption{Limits on $|d_t|$ at $\Lambda=1$ TeV applied to both same and opposite sign cases except for the cancellation region. The constraints for $\Lambda=10$ TeV are roughly $0.14$ times smaller.}
\begin{tabular}{c c}
\hline
Present~$(d_e,~d_n)$ & $|d_t|\lesssim 1.3\times 10^{-19}~e$ cm\\
Future ~$(d_e,~d_n)$ & $|d_t|\lesssim 1.5\times 10^{-20}~e$ cm\\
Future ~$(d_e,~d_n,~d_p)$ & $|d_t|\lesssim 6.4\times 10^{-21}~e $ cm\\
\hline
\end{tabular}
\label{talbe:limits_dt}
\caption{Limits on $|d_t|$ at $\Lambda=1$ TeV associated with the cancellation region. }
\begin{tabular}{c c}
\hline
Present~$(d_e,~d_n)$ & $|d_t|\lesssim 7.0\times 10^{-18}~e$ cm\\
Future ~$(d_e,~d_n)$ & $|d_t|\lesssim 1.0\times 10^{-19}~e$ cm\\
Future ~$(d_e,~d_n,~d_p)$ & $|d_t|\lesssim 5.0\times 10^{-20}~e$ cm\\
\hline
\end{tabular}
\label{talbe:limits_dt_loophole}
\end{table}
%
%
%
%
\section{Conclusion and Discussions}\label{sec:conclusion}
Due to its sizable Yukawa coupling, the top quark provides one of the most powerful windows into BSM physics. The top quark EDM is particularly interesting because it is sensitive to possible new sources of CPV and because one generally expects it to be enhanced relative to the light fermion EDMs by the ratio of the respective Yukawa couplings.  Above the EW scale $v$, the top EDM originates from two gauge-invariant operators, ${\cal O}_{tB}$ and ${\cal O}_{tW}$, that appear at the BSM scale $\Lambda$. These operators also induce light fermion EDMs at the two-loop level. Consequently, the stringent bounds on systems involving first generation fermion EDMs, including paramagnetic atoms and polar molecules, neutrons, and diamagnetic atoms, imply strong constraints on ${\cal O}_{tB}$ and ${\cal O}_{tW}$. By combining the results from these systems involving light fermions, one obtains tight bounds on $d_t$. The prospects for obtaining even greater sensitivity with future EDM experiments are promising.

The resulting present constraints and prospective sensitivities indicated in Tables \ref{talbe:limits_dt} and \ref{talbe:limits_dt_loophole} imply that $|d_t|$ is smaller than $\sim 10^{-19}$ $e$ cm, except in the presence of finely tuned cancellations between ${\cal O}_{tB}$ and ${\cal O}_{tW}$, allowing for a top EDM up to $\sim 50$ times larger. Next generation searches for the EDMs of the electron and neutron could yield up to a factor of ten increase in sensitivity, while a storage ring search for the proton EDM with sensitivity $|d_p|\sim 10^{-29}$ $e$ cm could lead to an additional  sensitivity increase. To the best of our knowledge, the $d_t$-reach of these experiments will exceed those of direct probes at the LHC.

Given these prospective sensitivities, it is important to bear in mind the opportunities for refined theoretical computations. In this work we have retained only the leading log-squared contribution to the RGE of ${\cal O}_{tB}$ and ${\cal O}_{tW}$ from  $\Lambda$ to $v$. The impact of subleading logarithmic contributions will be analyzed in a forthcoming publication \cite{KF_MJRM}. From the low-energy perspective, there exists room for refinements of the $d_n$ computations. While the uncertainties associated with the up- and down-quark EDMs enter at the $10\%$ level \cite{Bhattacharya:2015esa,Bhattacharya:2015wna}, those associated with the strange quark (not included in our study here) are considerably larger \cite{Bhattacharya:2015esa,Bhattacharya:2015wna}. In addition, BSM scenarios that induce ${\cal O}_{tB}$ and ${\cal O}_{tW}$ may also give rise to the corresponding CPV gluonic operators (CEDMs), a topic for which the phenomenology is considerably richer and the theoretical hadronic and nuclear uncertainties correspondingly more challenging. In that context, the interplay with LHC and future collider probes may be particularly enlightening.

\begin{acknowledgments}
We are grateful to Jordy de Vries and Adrian Signer for useful discussions and comments.
We also thank  Patrick Draper and Hiren Patel for having fruitful discussions.
KF thanks Natsumi Nagata and Eibun Senaha for valuable discussions. MJRM thanks Haipeng An, Mark B. Wise, and Yue Zhang for several helpful conversations. This work was supported in part under U.S. Department of Energy contract DE-SC0011095.
\end{acknowledgments}


\end{document}